\documentclass[pra,twocolumn,showpacs,preprintnumbers,amsmath,amssymb,floatfix]{
revtex4}
\usepackage{graphicx}
\usepackage{dcolumn}
\usepackage{bm}
\usepackage{latexsym,epsfig}

\newcommand{\beq}{\begin{equation}}
\newcommand{\eeq}{\end{equation}}
\newcommand{\bea}{\begin{eqnarray}}
\newcommand{\eea}{\end{eqnarray}}
\newcommand{\ben}{\begin{eqnarray*}}
\newcommand{\een}{\end{eqnarray*}}
\newcommand{\be}{\begin{enumerate}}
\newcommand{\ee}{\end{enumerate}}
\newcommand{\bfig}{\begin{figure}}
\newcommand{\efig}{\end{figure}}

\newcommand{\ba}{\begin{align}}
\newcommand{\ea}{\end{align}}

\newcommand{\D}{\displaystyle}

\newcommand{\la}{\langle}
\newcommand{\ra}{\rangle}

\newcommand{\vtrap}{V_{\text{T}}}

\begin{document}
\title{Supersolid in a one-dimensional optical lattice in the presence of a harmonic trap}
\author{Tapan Mishra}
\email{tapan@physics.georgetown.edu} \affiliation{ Indian Institute of
Astrophysics, II Block, Kormangala, Bangalore, 560 034, India. \\
Department of Physics, Georgetown University, Washington, DC 20057, USA}
\author{S. Ramanan}
\email{sramanan@ictp.it}
\affiliation{Centre for High Energy Physics,
Indian Institute of Science, Bangalore\ 560012, India, \\
The Abdus Salam International Center for Theoretical Physics, Trieste 34151,
Italy.}
\author{Ramesh V. Pai}
\email{rvpai@unigoa.ac.in} \affiliation{ Department of Physics, Goa
University, Taleigao Plateau, Goa 403 206, India. }
\author{Meetu Sethi Luthra}
\email{meetu@iiap.res.in} \altaffiliation [permanent address
]{Bhaskaracharya College of Applied Sciences, Phase-I,
Sector-2,Dwarka,Delhi,110075, India.} \affiliation{Indian Institute
of Astrophysics, II Block, Kormangala,  Bangalore,  560 034, India.}
\author{B. P. Das}
\email{das@iiap.res.in} \affiliation{Indian Institute of
Astrophysics, II Block, Koramangala, Bangalore, 560 034, India.}

\date{\today}

\begin{abstract}
We study a system of ultra-cold atoms possessing long range
interaction (e.g. dipole-dipole interaction) in a one dimensional
optical lattice in the presence of a confining harmonic trap. We
have shown that for large enough on-site and nearest neighbor
interaction a supersolid phase can be stabilized, consistent with
the previous Quantum Monte Carlo and DMRG results for the
homogeneous system. Due to the external harmonic trap potential the
supersolid phase coexists with other phases. We emphasize on the
experimental signatures of the various ground state phases in the presence of a
trap.
\end{abstract}


\maketitle

\section{INTRODUCTION}
\label{sect:intro}

The realization of the supersolid form of matter, where the
superfluid and the crystalline order co-exist~\cite{andreev,legget},
in the ultra-cold bosonic atoms in optical lattices is at the
forefront of research. After the claim of observing the supersolid
phase in solid $^4He$ by Kim {\it et al}~\cite{chan}, the progress
in the research of this exotic phase of matter has advanced
substantially. The successful observation of the superfluid (SF) to
Mott insulator (MI) transition in ultra-cold bosonic atoms in
$3D$~\cite{bloch} and subsequently in $2D$~\cite{spielman} and
$1D$~\cite{stoferle} has shaped the study of ultra-cold systems as an 
ideal tool to understand condensed matter phenomena. In order to achieve the
supersolid form of
matter that is characterized by the co-existence of the superfluid
and crystalline order, it is essential for the system to have long
range interactions. The remarkable experimental realization of BEC
in $Cr$ atoms~\cite{pfau} that have fairly large dipole moment, has
increased the expectations to observe the supersolid phase in
optical lattice experiments.

In recent years there have been several theoretical evidences for
the supersolid phase in various lattice
geometries~\cite{batrouniprl,mishrass,
kedar,arun,sengupta,wessel,scarola}. However, the experimental
search of the supersolid in the ultra-cold atomic systems in optical
lattices still remains a challenge. The real experimental situation
is different from the usual homogeneous system considered in
theoretical calculations. In experiments, the translational symmetry
of the lattice is broken due to the presence of an external harmonic
trap potential (magnetic or optical) and various quantum phases
co-exist
~\cite{wessel1,svistunov,bergkvist,pollet,smita,mitra,spiel,nandini,suna,batrounimi}.
Hence, it is essential to understand the signatures of the
supersolid phase in the presence of such a trap.

In this paper we have considered a system of ultra-cold bosonic
atoms possessing long range interactions in a one dimensional
optical lattice with a harmonic confinement. The Hamiltonian for
this kind of system is represented by the extended Bose-Hubbard
model,

\begin{eqnarray}
H &=&-t\sum_{<i,j>}(a_{i}^{\dagger}a_{j}+H.c)
+\frac{U}{2}\sum_{i} n_{i}(n_{i}-1)\nonumber\\
& &\mbox{} +V\sum_{<i,j>}n_in_j+\vtrap\sum_i\,r_i^2n_i.
\label{eq:ham}
\end{eqnarray}

Here, $t$ is the hopping amplitude between the nearest neighbor
sites $\la i,j \ra$, $a_i^{\dagger}(a_i)$ is the bosonic creation
(annihilation) operator obeying the Bosonic commutation relation
$[a_i,a_j^\dagger] = \delta_{i,j}$ and $n_i = a_i^{\dagger}a_i$ is
the number operator. $U$ and $V$ are the on-site and the nearest
neighbor interactions, respectively. $\vtrap$ is the magnitude of
the external trap potential and $r_i$ is the distance from the trap
center. We re-scale in units of the hopping amplitude, $t$, setting
$t= 1$, making the Hamiltonian and other quantities dimensionless.

The homogeneous version of this model (i.e. without the external
trap), has been studied earlier using several techniques in one
dimension~\cite{mishrass,pai,whiteprb,batrouniprl,kashurnikov,
batrouni95,niyaz,kuhner,iskin}. The prediction of an accurate phase
diagram using Quantum Monte Carlo method~\cite{batrouniprl} and
DMRG~\cite{mishrass} has revealed the physical conditions required
to stabilize a supersolid phase. It has been shown that the
supersolid phase is obtained when:
\begin{enumerate}
\item The total density of the system is incommensurate to the lattice.
\item The on-site ($U$) and the nearest neighbor interactions ($V$) are
fairly large compared to the hopping amplitude ($t$).
\item The condition $U<2V$ is satisfied.
\end{enumerate}
A homogeneous system exhibits a uniform phase determined by the
global chemical potential for a given set of interaction parameters.
The phase diagram of the model in Eq.~\ref{eq:ham} in the homogeneous
limit i.e., $\vtrap=0$, exhibiting different possible phases
including the supersolid phase is shown in
Fig.~\ref{fig:fig1}~\cite{mishrass}.
\bfig[ht]
  \centering
\includegraphics[width = 3.4in, angle = 0, clip = true]
{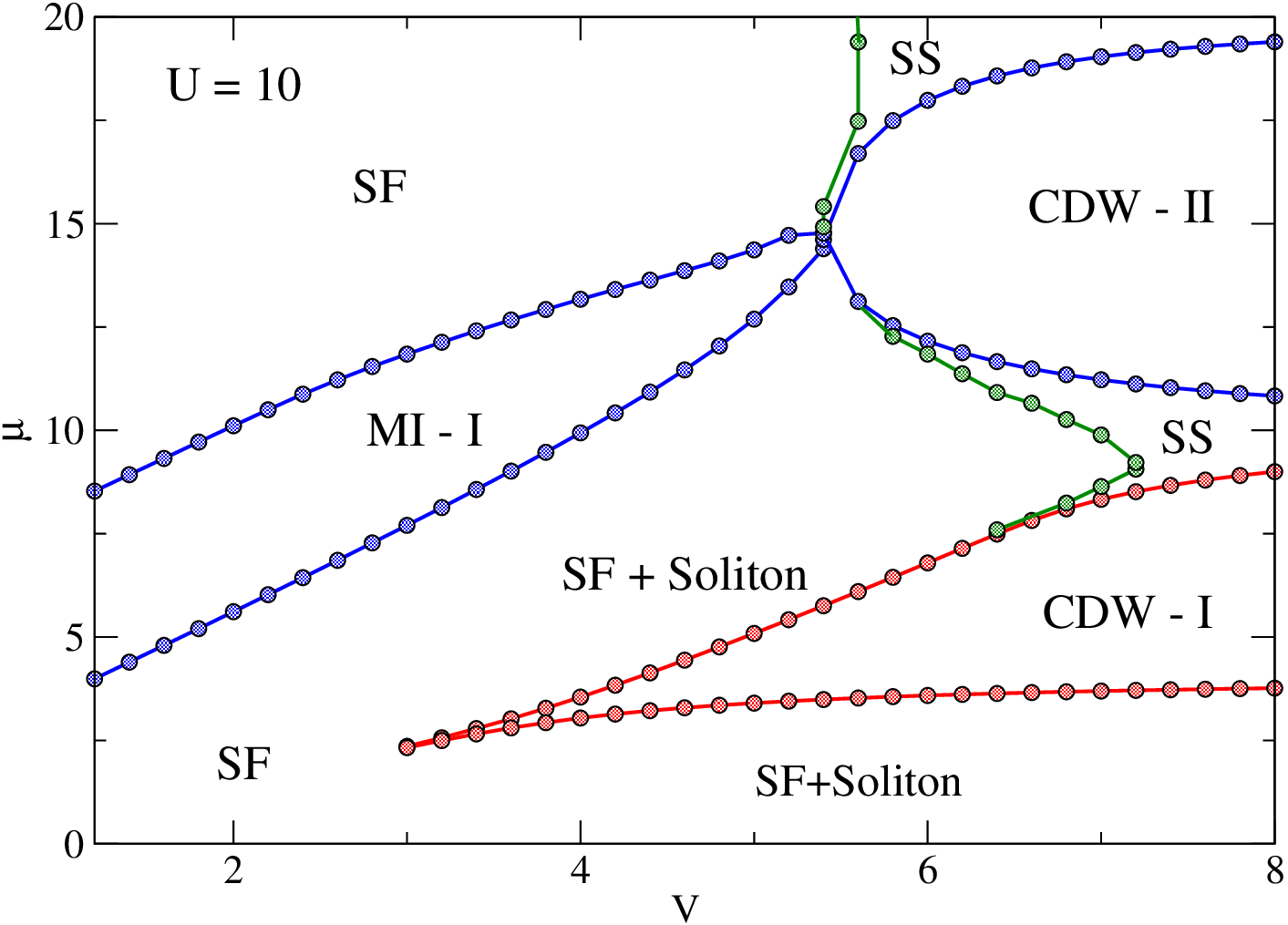}
    \caption{(Color on line) The phase diagram of the model in
Eq.~\ref{eq:ham}
in the homogeneous limit i.e., $\vtrap=0$ and for $U= 10.0$ in the
$\mu-V$ plane.}
    \label{fig:fig1}
  \begin{center}
   \includegraphics[angle = 0, width = 3.4in, clip = true]
{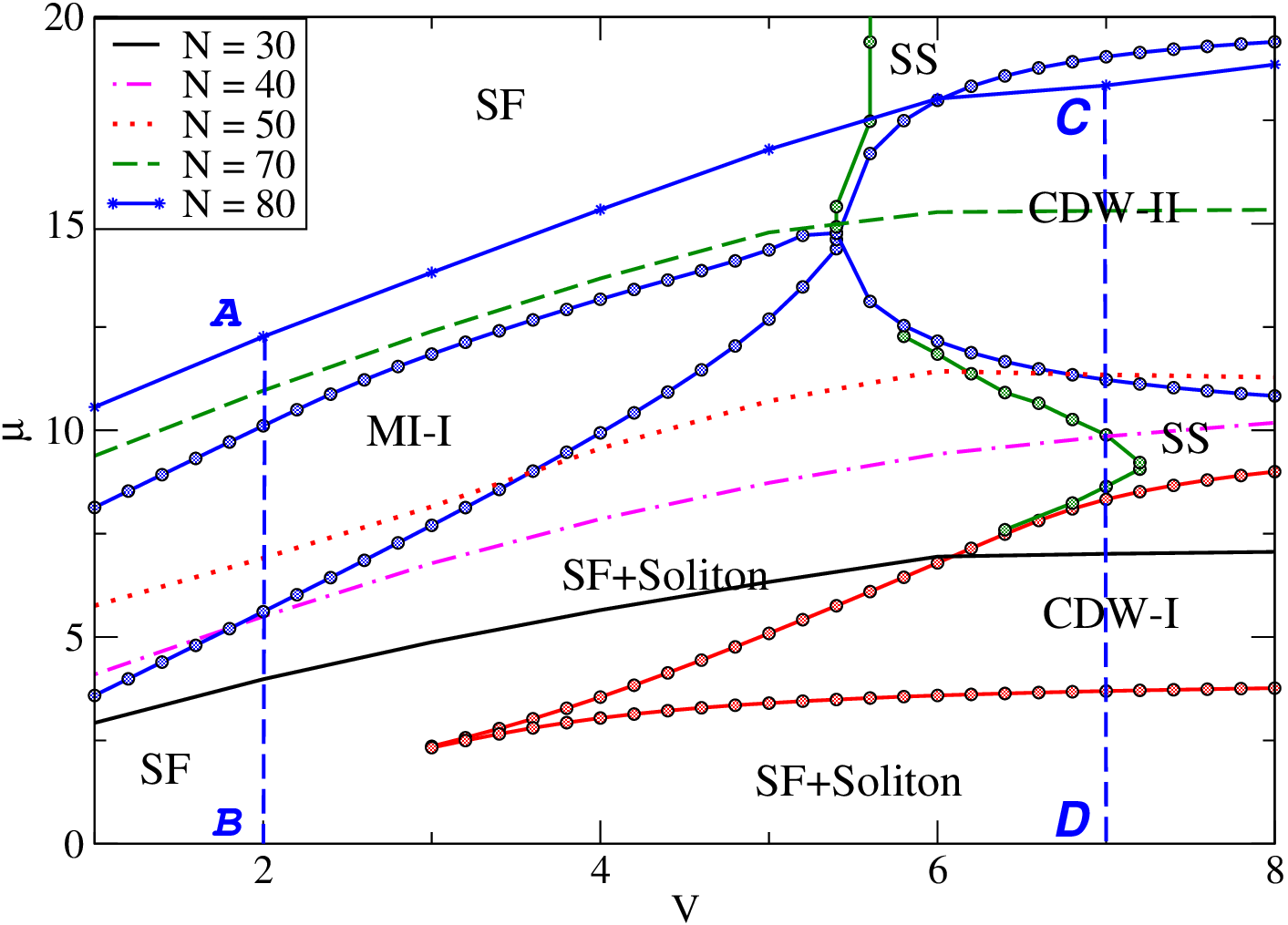}
  \end{center}
\caption{(Color on line) Homogeneous phase diagram for the
model in Eq.~\ref{eq:ham} showing canonical trajectories. For a given
value of $V$, the presence of an external trap allows all phases
that fall on the line that starts from the canonical trajectory to
the V-axis for example: lines AB and CD in the figure.}
  \label{fig:fig1a}
\efig
In the presence of an external trap, the role of a local chemical
potential becomes important as demonstrated in our earlier work
on the Bose-Hubbard model~\cite{suna}. Since the local chemical
potential varies from the center of the trap to the edges, the system
exhibits different phases simultaneously. An earlier DMRG study of
model given in Eq.~\ref{eq:ham} could not confirm the presence of
the supersolid phase in the system~\cite{urba}. A recent study of
this model in two dimension using mean field theory predicts that
the noise correlation could be a valid signature to separate the
supersolid phase from the other ground state phases~\cite{scarola}. In this
paper we re-visit the extended Bose-Hubbard model with the external harmonic
trap potential and search for experimental signatures of the different
ground state phases, in particular, the supersolid phase.

The remaining part of the paper is organized as follows. In
Sec.~\ref{sect:meth}, we discuss the method of our calculation using the
finite size density matrix renormalization group (FS-DMRG)
technique. The results along with discussions are presented in
Sec.~\ref{sect:res_disc} with experimental signatures for the
different ground state phases in Sec.~\ref{sect:exp_sig} and we present
our conclusions in Sec.~\ref{sect:concl}.

\section{METHOD OF CALCULATION}
\label{sect:meth}

To obtain the ground state of model~(\ref{eq:ham}) for the system of
$N$ bosons on a lattice of length $L$, we use the FS-DMRG method
with open boundary conditions~\cite{white,schollwock}. This method
has been widely used to study the Bose-Hubbard
model~\cite{pai,kuhner,whiteprb,schollwock,urba}. We have considered
six bosonic states per site and the weights of the states neglected
in the density matrix formed for the left or the right blocks are
less than $10^{-6}$~\cite{pai}. In order to improve the convergence
of the results, the finite-size sweeping procedure as given
in~\cite{white,pai} has been used for every length. Using the ground
state wave function $|\psi_{LN} \ra$ and energy $E_L(N)$, we
calculate the following physical quantities and use them to identify
the different phases.

The on-site local number density $\la n_i \ra$, defined as,
\beq
    \la n_i \ra= \la\psi_{LN}|n_i|\psi_{LN} \ra,
    \label{eq:ni}
\eeq gives the local density distribution.
The fluctuation in the local number density, $\kappa_i$, which is finite for
the SF phase, is calculated
using the relation \beq
    \kappa_i =\la{n_i^2}\ra -{\la{n_i}\ra}^2
\label{eq:kappa} \eeq
 and finally the existence of the CDW order is confirmed by
calculating the structure factor:
\beq
 S(k)=\frac{1}{L^2}\sum_{i,j}e^{i\,k\,(i-j)}\la{n_in_j}\ra.
\label{eq:sk}
\eeq
In our calculations, we have considered a system of length $L=140$ and vary
$N$ from $30$ to $140$. In our previous work on the homogeneous
extended Bose-Hubbard Model~\cite{mishrass}, we had considered a
fixed value of the on-site interaction $U=10$ and vary the nearest
neighbor interaction strengths $V$ from $0$ to $10$. We choose the
same range of parameters here as well, since the homogeneous phase
diagram for this range, as shown in the Fig.~\ref{fig:fig1}, exhibits most of
the interesting phases for this model. The strength of the external
confining trap potential is fixed at $\vtrap=0.008$.

\section{Results and Discussion}
\label{sect:res_disc} We begin with the summary of the phase diagram
for the homogeneous extended Bose-Hubbard model, which has been
studied recently~\cite{batrouniprl,mishrass} for a wide range of
densities and interaction parameters namely, the on-site
interaction $U$ and the nearest neighbor interaction $V$. The phase
diagram for a typical value of the on-site interaction, say $U=10$ is
shown in Fig.~\ref{fig:fig1}~\cite{mishrass}. The phase diagram consists of
gapped as well as gapless phases. The gapless phases include the superfluid
phase, the supersolid phase where superfluidity and charge
density wave order co-exist, and the solitonic phases. The gapped
phases are (i) the Mott insulator phase with $\rho=1$ for $V < V_C
\sim 5.4$, (ii) charge density wave phase
CDW-II, (where every other site is doubly occupied, i.e.,
$|2~0~2~0~\cdots\ra$) with average density $\rho=1$ for $V> V_C\sim
5.4$ and (iii) the CDW-I phase (alternative sites are occupied, i.e., boson
density varies as $|1~0~1~0~\cdots\ra$) with average density
$\rho=1/2$ for $V > V_C \sim 3.0$. The gap vanishes
when doping above or below these gapped phases.
For example doping below half-filling ($\rho=1/2$) gives rise to
solitons that break the CDW-I order. This phase extends over a small
range of densities below the CDW-I and eventually goes over to the
superfluid phase when the density is further decreased. However, the
behavior of the system when doping above half-filling is different. For small
$V$ we get similar solitonic phases, however, for larger $V$ a
supersolid phase stablizes. The supersolid phase forms again while
doping above and below the CDW-II phase. In fact there exists a range
densities $0.5<\rho<1$ and $\rho>1$ for $V
> U/2$ where the supersolid phase is the stable ground state of
model~(\ref{eq:ham}) as shown in the Fig.~\ref{fig:fig1}.

Let us introduce a harmonic trap potential. Earlier studies of the
one-dimensional Bose-Hubbard model in the presence of a trap has
demonstrated the co-existence of the superfluid and the Mott
insulator phases~\cite{suna,batrounimi}. The Mott insulator is
characterized by the formation of a plateau in the local number
density $\la n_i \ra$ as a function of the distance $r_i$ from the
center of the trap and is incompressible, while the superfluid phase
is characterized by large local number density fluctuations and is
compressible. The nearest neighbor interaction brings about the
charge density wave order in the system due to the interplay between
the $U$ and $V$ terms in the Hamiltonian. Figures~\ref{fig:fig2}
and~\ref{fig:fig2a} show the density profile, i.e., the variation of
the local density $\la n_i\ra$ as a function of the distance from
the trap center $r_i$. We obtain the density profile for two sets of
parameters: (i) for the number of bosons fixed at $N=80$, but
different values of $V$ (Fig.~\ref{fig:fig2}) and (ii) fixed nearest
neighbor interaction, $V=8$, but different values of $N$
(Fig.~\ref{fig:fig2a}). The following three features are clearly
seen: (i) the local density $\la n_i\ra$ is maximum at the center of
the trap, (ii) the density falls-off with increase in $r_i$ and
(iii) the density profile exhibits plateaus and oscillations.

In order to understand these features and identify various phases from the
density profile, we define the local chemical potential at
the site $i$ at a distance $r_i$ from the center of the trap as,
\beq
  \mu_i = \mu_0 - \vtrap r_i^2.
  \label{eq:local_mu}
\eeq
Here $\mu_0=E_L(N+1)-E_L(N)$ is the chemical potential of the
system. For the homogeneous system, $\mu_i=\mu_0$ for any $i$.
However, for a finite trap the local chemical
potential $\mu_i$ equals $\mu_0$, which is its maximum value, at the center of
the trap and decreases radially outward as in Eq.~\ref{eq:local_mu}. It is
instructive to plot the density profile as a function of $\mu_i$ instead of
$r_i$ as in Fig.~\ref{fig:ni_mui_3}.
It may be noted from Fig.~\ref{fig:ni_mui_3} that the density of bosons at
any site $i$ is controlled by the value of the local chemical potential
$\mu_i$. So a decrease in $\mu_i$ results in a decrease in $\la n_i
\ra$, with the maximum at the center of the trap as observed in
Figs.~\ref{fig:fig2} and~\ref{fig:fig2a}.

 \bfig[ht]
  \centering
\includegraphics[width = 3.4in, angle = 0, clip = true]
{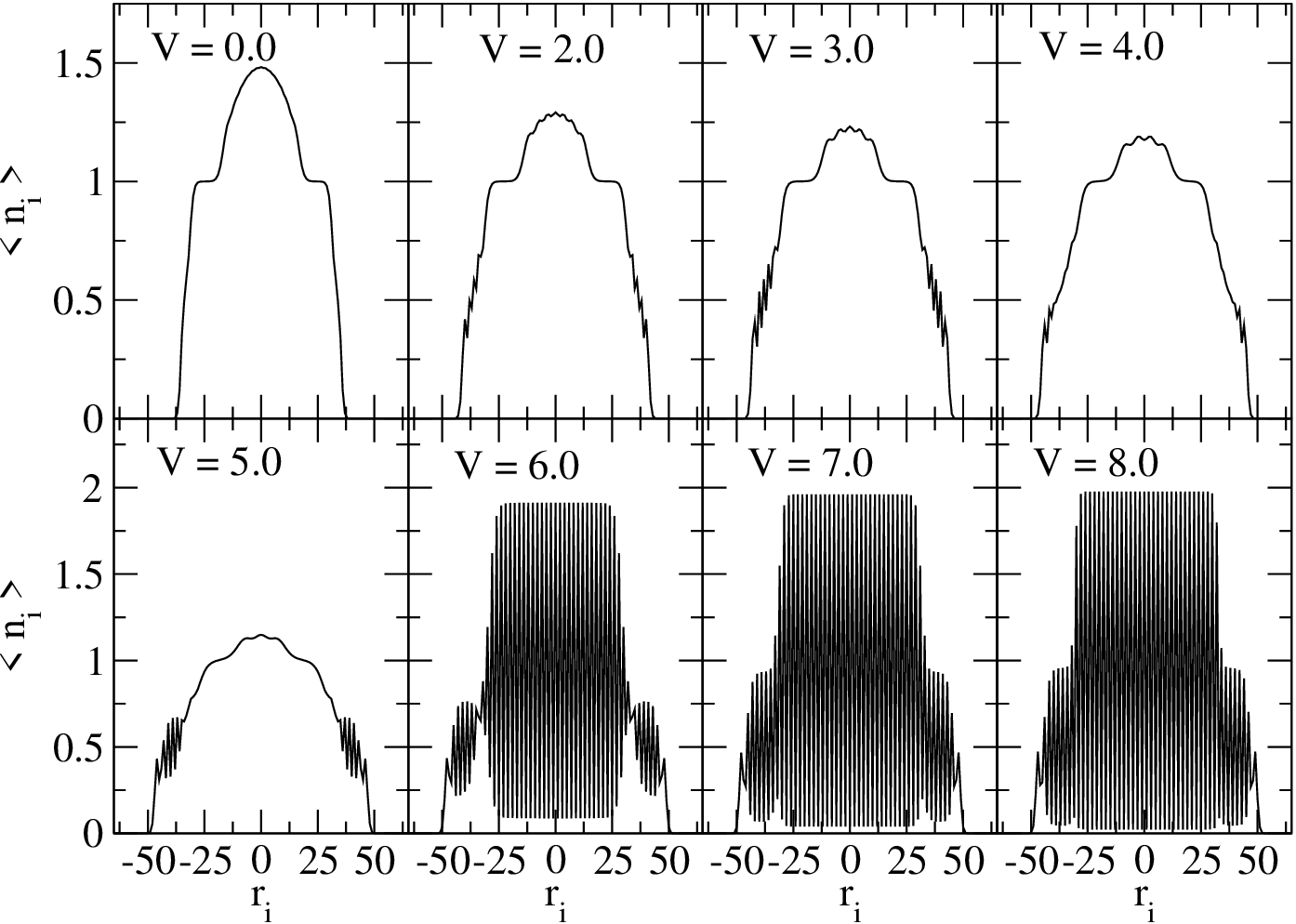}
    \caption{The local density $\la n_i \ra$ as a function of
the distance from the center of the trap $r_i$
    for $N = 80$, $U=10$, $\vtrap=0.008$, but for different values of $V$.
    }
    \label{fig:fig2}
\efig

\bfig[ht]
  \centering
\includegraphics[width = 3.4in, angle = 0, clip = true]
{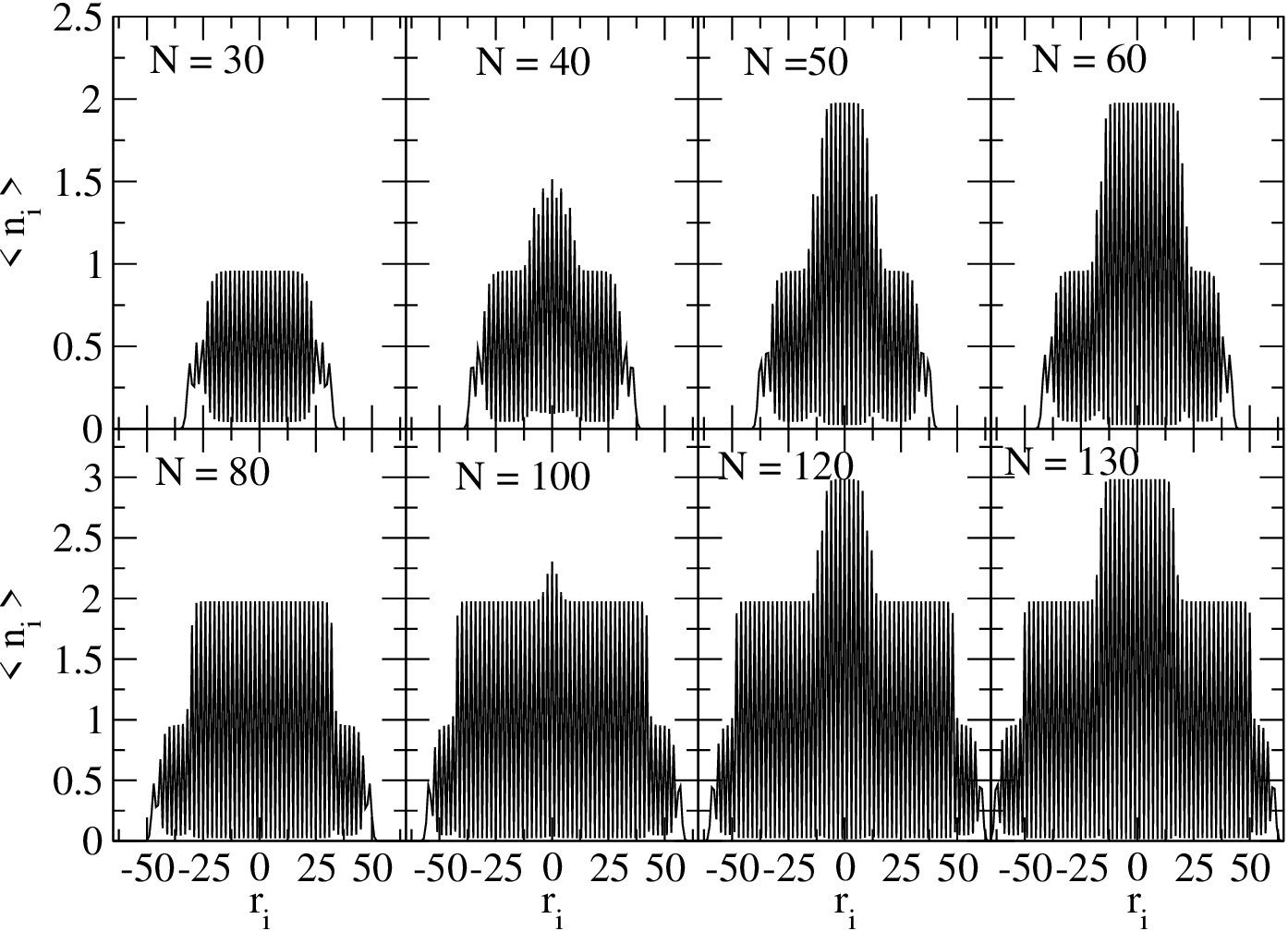}
    \caption{The local density $\la n_i \ra$ as a function of the distance from
the center of the trap $r_i$
    for fixed $V = 8$, $U=10$, $\vtrap=0.008$, but for different values of
    $N$.
    }
    \label{fig:fig2a}
\efig

The density of bosons plays a very crucial role in the determination
of the ground state of the model in Eq.~\ref{eq:ham}. The gapped
phases are possible only when the density is commensurate. The
homogeneous system with a given value of $U$ and $V$ and a uniform
local chemical potential $\mu_0$ represents one point in the phase
diagram. However, for the system with a trap potential, the density
varies across the lattice due to the variation of the local chemical
potential and therefore different phases co-exist. In order to
understand this feature of co-existence of the different ground
state phases and the role played by the local chemical potential, we
study first the path in the phase diagram that is traced by $\mu_0$
as we change the interaction parameter $V$ keeping the number of
bosons $N$, the trap potential $V_T$ and the on-site interaction $U$
fixed. This path is referred to as the \emph{Canonical
Trajectory}~\cite{batrounimi}, since $N$ is held fixed.
Fig.~\ref{fig:fig1a} shows several canonical trajectories (for
different values of $N$) in the homogeneous phase diagram. In may be
noted that $\mu_0$ is the local chemical potential at the center of
the trap and the position of these canonical trajectories trace the
phase present at the trap center as $V$ is varied for fixed $N$. For
example when $N=30$, the canonical trajectory and hence the phase at
the center of the trap goes from the superfluid to CDW-I as we
increase $V$. The position of the canonical trajectory in the phase
diagram can be shifted by changing the number of bosons $N$. When
the number of bosons is increased, say to $N=40$, $\mu_0$ increases
and the position of the canonical trajectory in the phase diagram is
shifted upward. As a result the center of the trap, say for $V=0$,
which was in the SF phase for $N=30$, is now in the Mott insulator
phase. Following the canonical trajectory for $N=40$, the trap
center goes from MI to SF and then to a supersolid phase for
increasing $V$. Thus the position of the canonical trajectory for a
given $N$ and $V$ in the phase diagram represents the phase at the
center of the trap.

Moving away from the center of the trap, the local chemical
potential decreases as in Eq.~(\ref{eq:local_mu}) and the
variation of $\mu_i$ is represented in the phase diagram by a line
drawn vertically downwards from the canonical trajectory to the
horizontal axis. The local chemical potential values across the
lattice fall on this line, which passes through different ground state
phases. Therefore, the local chemical potential (and thus local density) at
different sites favor the co-existence of different phases in the presence of
a trap. It is useful to re-plot the density profile given in the
Fig.~\ref{fig:fig2} as a function of $\mu_i$ using
Eq.~\ref{eq:local_mu} instead of $r_i$ as in
Fig.~\ref{fig:ni_mui_3}. We also calculate and plot, in the same
figure, the average local number density define as 
\beq
	\bar{n}_i= \la (2n_i+n_{i+1}+n_{i-1}) \ra/4.
	\label{eq:nbar}
\eeq

For $N = 80$ and $V = 2.0$, $\mu_0$ falls in the superfluid phase
above the $\rho=1$ Mott lobe (point $A$ as indicated on the
canonical trajectory corresponding to $N= 80$ in
Fig~\ref{fig:fig1a}). This means that the center of the trap has
$\la n_i \ra > 1$. Moving away from the trap center, $\mu_i$
decreases along the line $AB$ and there are regions where $\mu_i$
falls inside the MI lobe. From Fig.~\ref{fig:ni_mui_3}, we see that
for these values of $\mu_i$, $\la n_i \ra=1$. Similarly as we move
towards the edge, the values of $\mu_i$ decreases further such that the system
is once again in a superfluid phase on the lower side of the Mott lobe. So the
system for $N=80$, $V=2$ has a superfluid core
flanked by a MI phase and finally ending with a superfluid edge. The density
profile (top panel of Fig~\ref{fig:fig2} and
Fig~\ref{fig:ni_mui_3}) correlates with this result. In addition,
there are oscillations in $\la n_i \ra$ in the superfluid shoulders
near $\bar{n}_i=1/2$. The reasons for these oscillation are the
following. For $V=2$, the system is close to CDW-I lobe (see
Fig~\ref{fig:fig1}). In the thermodynamic limit, the CDW-I order can stabilize
only for $V> V_C\sim 3.0$. However, the finite size of the system allows a
CDW-I phase to exist for lower values of $V$, here $V=2$, although it
vanishes in the thermodynamic limit. Finite size effects are characterized by
oscillations in the local density $\la n_i \ra$. These oscillations
stabilize at higher values of $V$ into the CDW-I phase. For example for $V=7$
and $N=80$, $\mu_0$ (point $C$ in Fig.~\ref{fig:fig1a}) falls inside the CDW-II
lobe yielding a CDW-II phase at the center. As we move towards
the edges, the CDW-II phase is flanked by a supersolid phase,
CDW-I and finally a superfluid shoulder as can also be infered from the density
profiles as shown in Fig.~\ref{fig:fig2} and Fig~\ref{fig:ni_mui_3}. In
fact, these conclusions can be further fortified by comparing the variation
of the average $ \bar{n}_i$ as a function of $\mu_i$ with the density of the
corresponding homogeneous system as in Fig.~\ref{fig:ni_mui_3a}. The
agreement is striking, leading to the conclusion that for a given set of
parameters, the phase of a system with an external trap
is represented by a line starting from the canonical trajectory to the
horizontal axis while the phase of the homogeneous system is represented by a
point in the phase diagram. This immediately shows that while the homogeneous
system can have a unique phase, the phases tend to co-exist for an
inhomogeneous system.

In addition to scanning along the phase diagram at fixed values of
$N$ and varying $V$, it is also equally possible to fix the nearest
neighbor interaction $V$ and move along the phase diagram by varying
$N$ and therefore the chemical potential $\mu_0$. The canonical
trajectory in the phase diagram moves upwards (downwards) by
increasing (decreasing) the total number of bosons and as a result
the local chemical potential at the center of the trap $\mu_0$
changes, giving rise to different phases at the center. This is
demonstrated in Fig.~\ref{fig:fig2a} for fixed $V=8$ for different
values of $N$. For $N=30$, position of the $\mu_0$ is inside the
CDW-I lobe (see Fig.~\ref{fig:fig1a}) and as discussed above, the
corresponding system has a CDW-I core flanked by a superfluid edge
as seen in the density profile (top panel of Fig.~\ref{fig:fig2a}). An
interesting situation occurs for
$N = 40$, where the trap center is expected to be in the elusive
supersolid phase, as seen in Fig.~\ref{fig:fig1a} and is
characterized by density fluctuations between $1.0 \le \la n_i \ra
\le 1.5$, that is, the system has a CDW order at incommensurate
densities~\cite{mishrass}. As a result, the system now will have a
supersolid core, followed by a CDW-I and a superfluid phase moving
outward from the trap center (top panel of
Fig.~\ref{fig:fig2a}). Further increase in $N$ leads to the inclusion
of a CDW-II phase in the system in addition to the supersolid, the
CDW-I and the superfluid phases.

\bfig[ht]
  \begin{center}
   \includegraphics[angle = 0, width = 3.4in, clip = true]{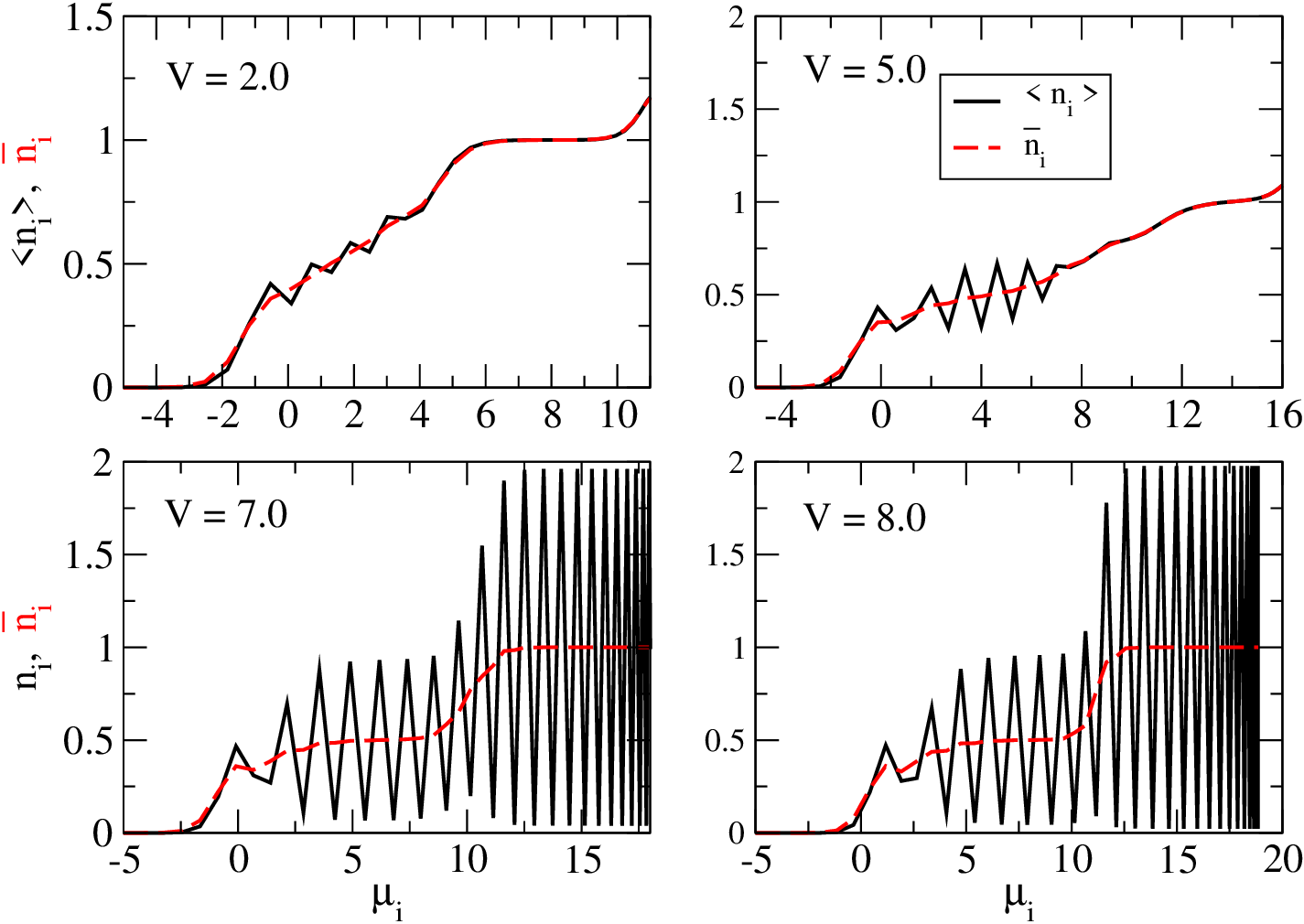}
  \end{center}
  \caption{(Color on-line) Local number density $\la n_i \ra$ and average define as
  $\bar{n}_i= \la (2n_i+n_{i+1}+n_{i-1})/4 \ra$ as a function of local chemical potential
for different values of $V$ but
  fixed $N = 80$.}
  \label{fig:ni_mui_3}
\efig \bfig[ht]
  \centering
\includegraphics[width = 2in, angle = 0, clip = true]
{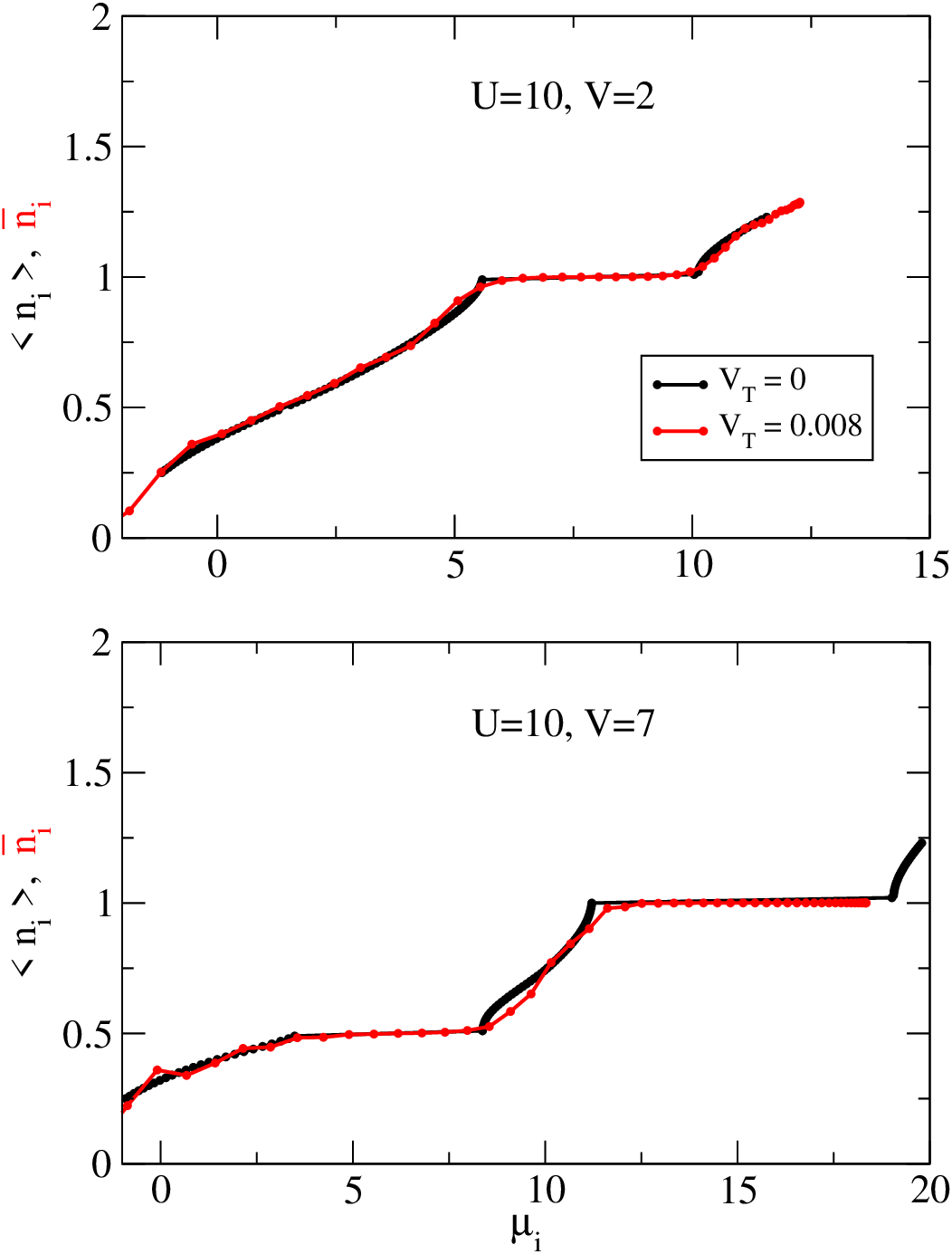}
    \caption{(Color on-line) Average local number density $\bar{n}_i$ for system
with a trap and $\la n_i \ra$ for a homogeneous system as function of the
    local chemical potential $\mu_i$ for $V=2$ and $7$.}
    \label{fig:ni_mui_3a}
\efig \bfig[ht]
  \begin{center}
   \includegraphics[angle = 0, width = 3.4in, clip = true]{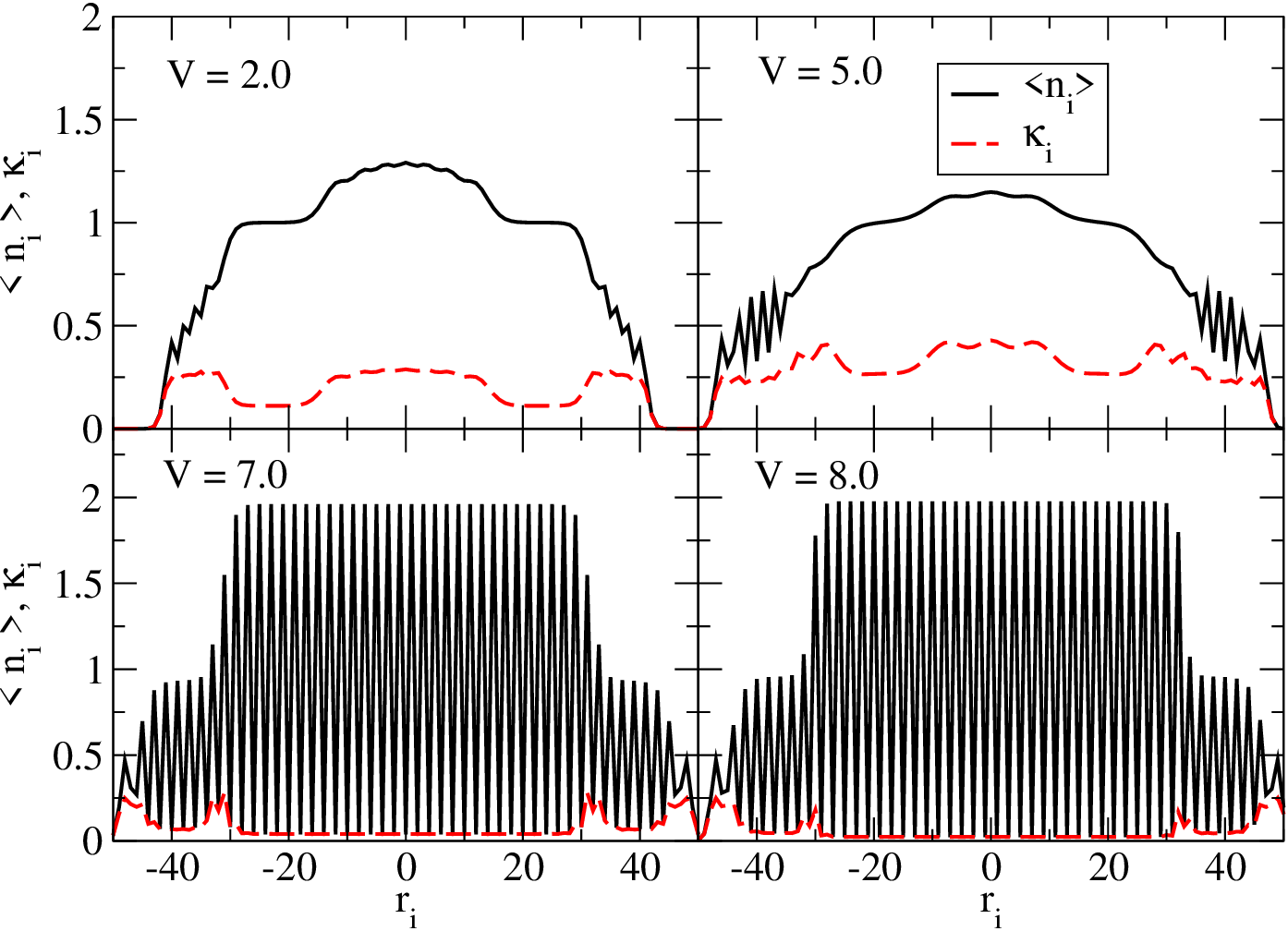}
  \end{center}
  \caption{(Color on-line) Number density per site and its fluctuations that serve as a measure of compressibility and
  hence can be used as a tool to pick out the compressible and the incompressible phases that coexist in the presence of a harmonic trap.}
\label{fig:fig4}
\efig

\bfig[ht]
  \begin{center}
   \includegraphics[angle = 0, width = 3.4in, clip = true]{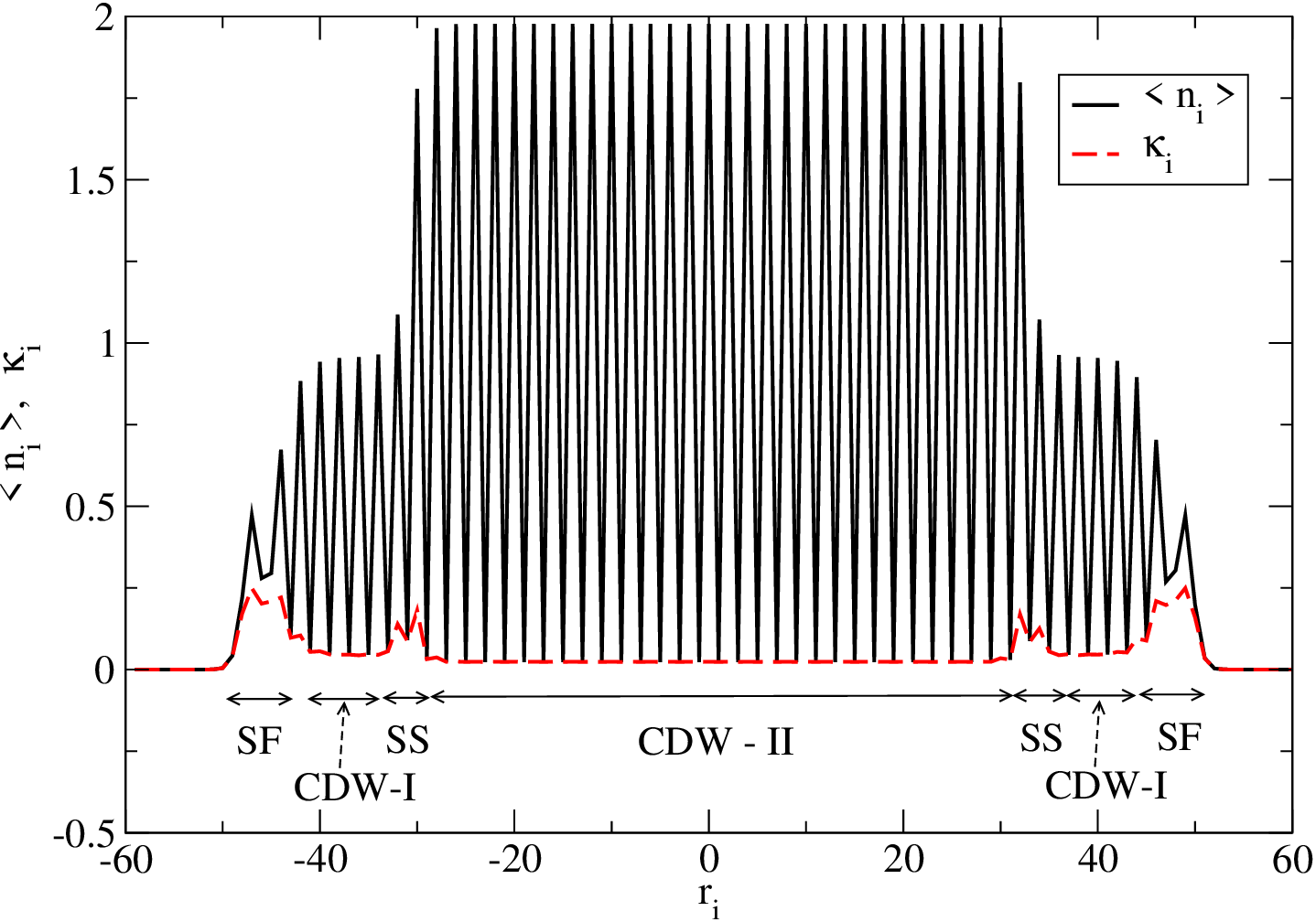}
  \end{center}
  \caption{(Color on-line) Picking out the different phases from the density profile for $N = 80$ and $V = 8.0$.}
  \label{fig:fig5}
\efig

The next issue we address here is a scheme to pick out the various
phases using local properties of the system.  We will follow the
discussions in~\cite{suna} and use local compressibility or
equivalently the fluctuations in the number density per lattice site,
$\kappa_i$, given in Eq.~\ref{eq:kappa}, as a tool to distinguish
between the gapped and the gapless phases. It is known that the
number fluctuation is large in the superfluid phase while it is a
minimum for the MI and the CDW phases. Fig.~\ref{fig:fig4} shows the
variation of $\kappa_i$ across the lattice. For small values of $V$,
$\kappa_i$ varies at the center and at the edges of the trap indicating that
these regions are in the superfluid phase, while the plateaus represent the
Mott insulator phase. Further, we note
that these plateaus (minima) occur exactly over the values of $r_i$ where the
average local density $\bar{n}_i$ exhibits a plateau at integer densities.
Therefore, one can pick out the incompressible phases using the
density profile and its local fluctuation $\kappa_i$ and identify
them using the phase diagram and the canonical trajectories.
As an example, Fig.~\ref{fig:fig5} shows the different phases for $N=80
$ but varying $V$. In the next section, we will discuss the
experimental signatures for the various phases that have been
isolated in the presence of a harmonic trap using global properties
of the system.

\section{Experimental Signatures}
\label{sect:exp_sig}

The presence of a harmonic trap in the optical lattice leads to the
co-existence of the superfluid, the Mott insulator, the charge density wave and
the supersolid phases as seen in the previous sections. As a result, extracting
the signature of a particular phase in the presence of other phases becomes a
theoretically important exercise in order to make connections with experiments.
In the following we analyze possible global signatures of the
various ground state phases that can be experimentally confirmed.

It is now possible in experiments to record the spatial distribution
of the lattice with different filling
factors~\cite{bloch06,gemelke,sherson,bakr}.
Similar experiments in one-dimensional
optical lattices can yield density profiles using which the ground state phases
can be mapped. Another way to obtain direct
information about the Mott plateaus (shells in 3D) is through the
atomic clock shift experiment~\cite{campbel}. By using density
dependent transition frequency shifts, sites with different
occupation can be spectroscopically distinguished, thus giving us
information about the number of sites corresponding to a given density
$\rho$ of bosons, defined as $N(\rho)$. As a first step, we look for the
signatures of the solid phases (MI and CDW) in an atomic clock shift
experiment.

\bfig[ht]
  \begin{center}
   \includegraphics[angle = 0, width = 3.4in, clip = true]{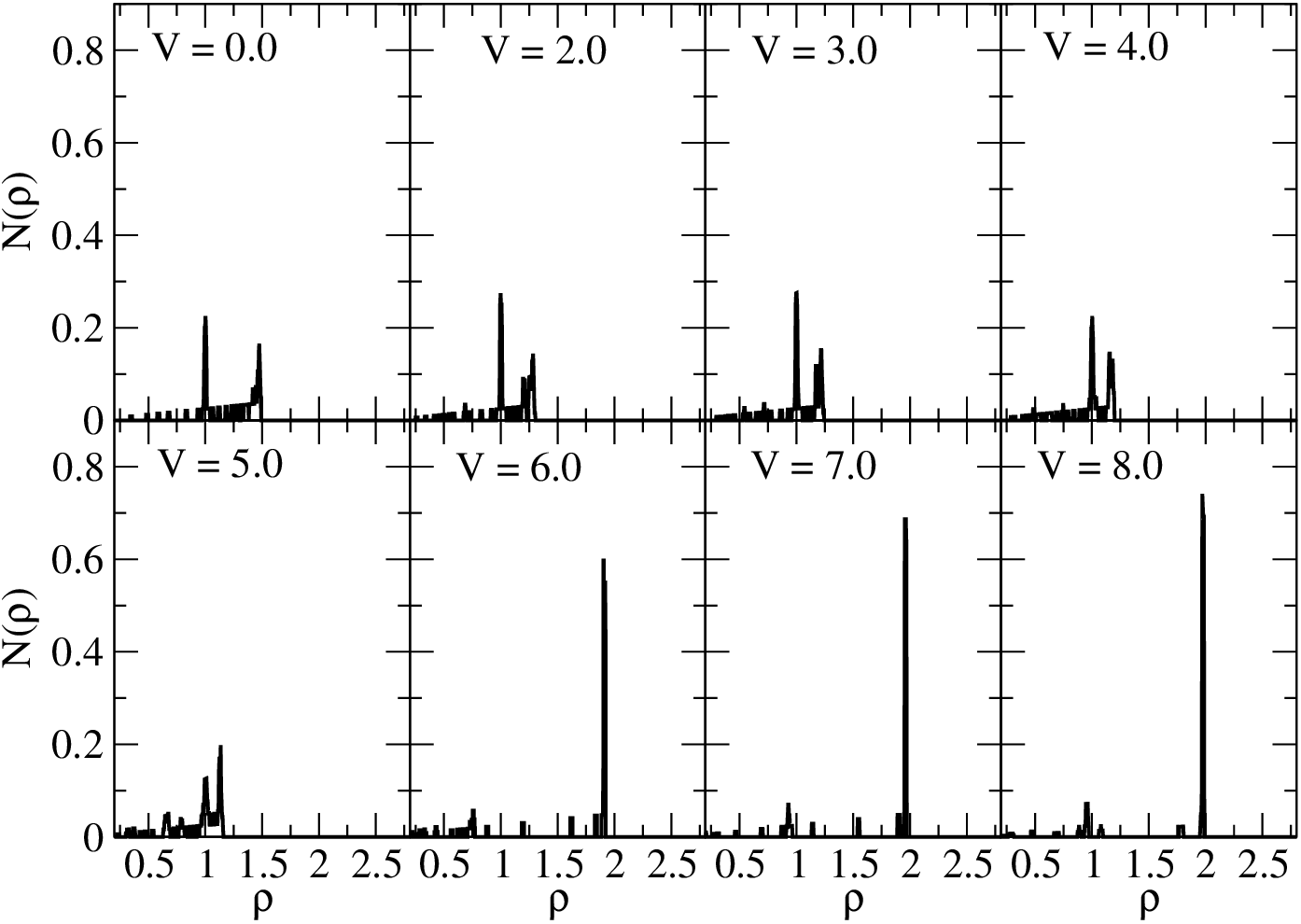}
  \end{center}
  \caption{$N(\rho)$ versus $\rho$ for $N = 80$ and different values of the
nearest neighbor interaction $V$. The
  presence of incompressible phases can be distinguished by the formation of a peak at commensurate densities.}
  \label{fig:fig6}
  \begin{center}
   \includegraphics[angle = 0, width = 3.4in, clip = true]{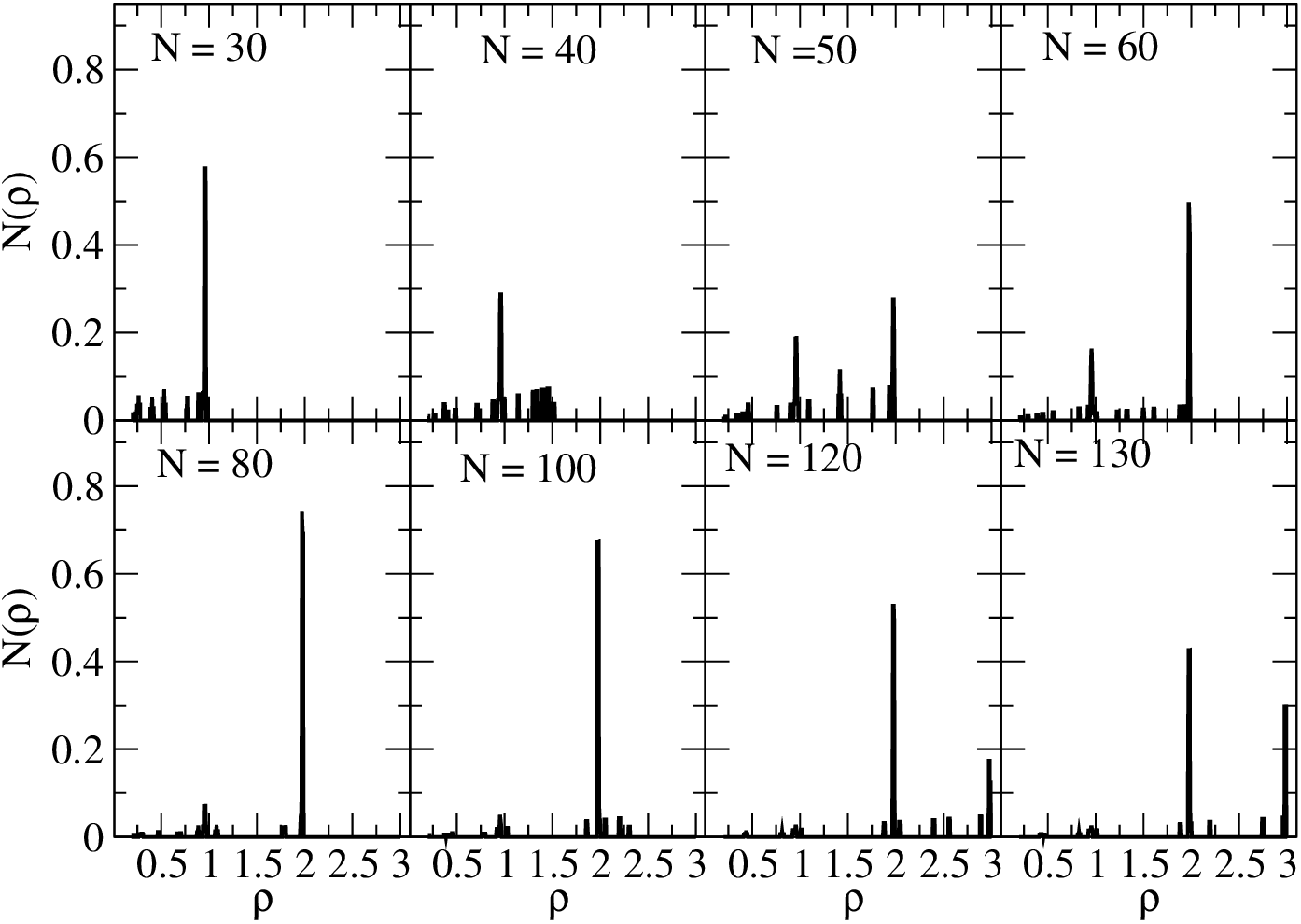}
  \end{center}
  \caption{$N(\rho)$ versus $\rho$ for $V = 8.0$ as number of Bosons is varied. Incompressible phases can be
  picked out by the formation of peaks at commensurate densities.}
  \label{fig:fig6a}
\efig

\bfig[ht]
  \begin{center}
   \includegraphics[angle = 0, width = 3.4in, clip = true]{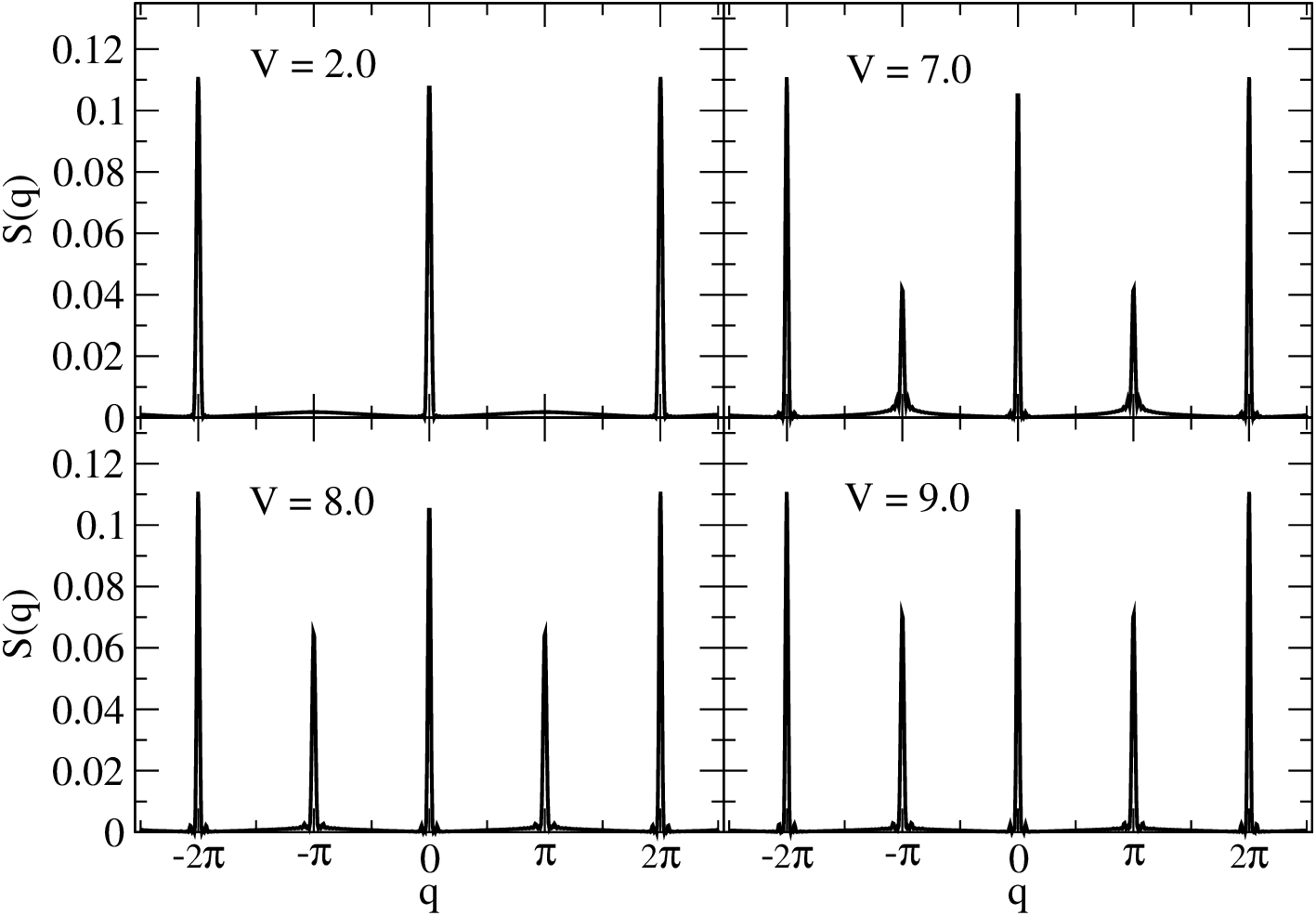}
  \end{center}
  \caption{Structure factor as a function of $q$ for $N = 40$ and different
values of $V$.}
  \label{fig:fig7}
  \begin{center}
   \includegraphics[angle = 0, width = 3.4in, clip = true]{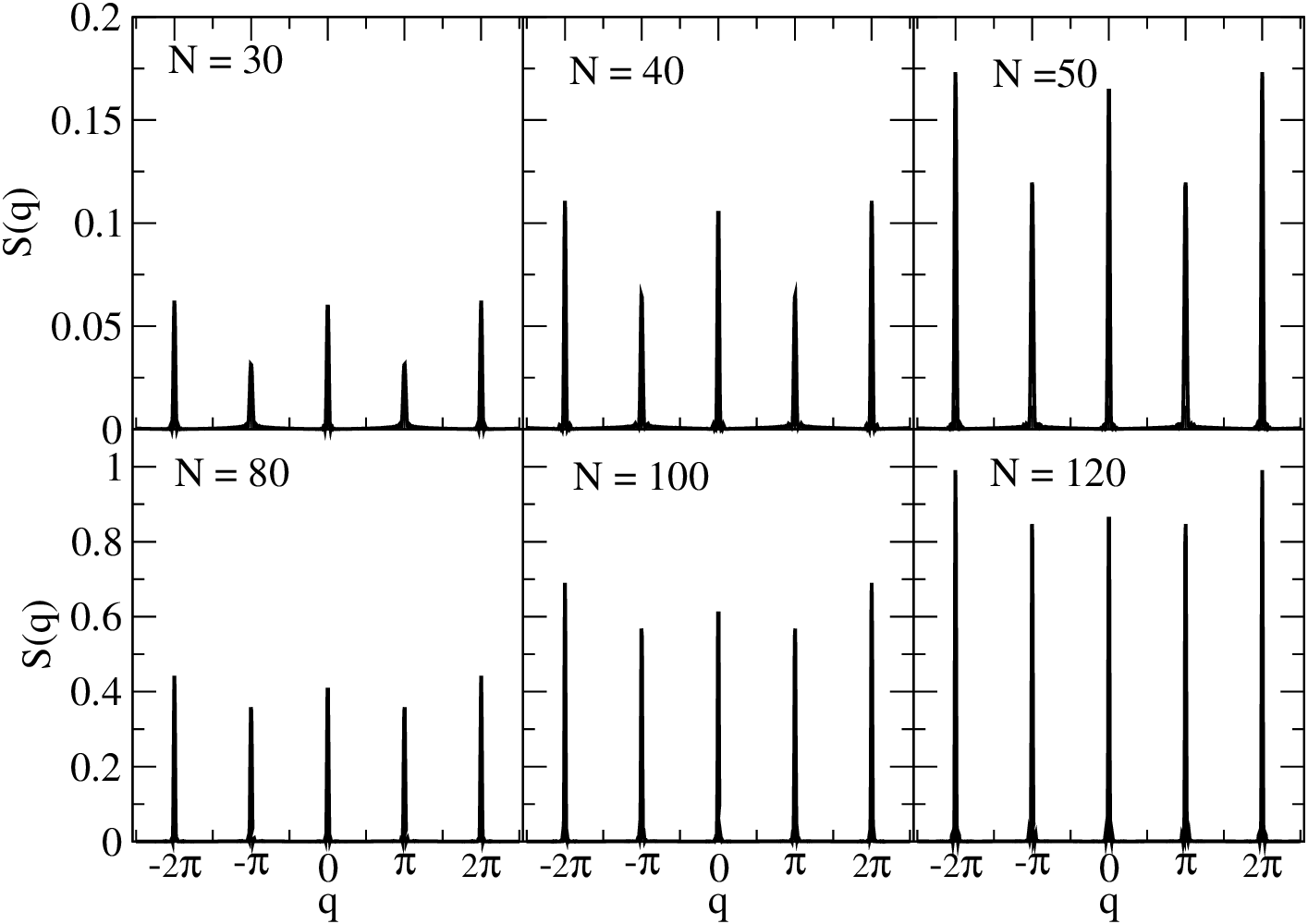}
  \end{center}
  \caption{Structure factor as a function of $q$ for $V = 8.0$ and different
values of $N$.}
  \label{fig:fig7a}
\efig

In Figs.~\ref{fig:fig6} and~\ref{fig:fig6a} we plot $N(\rho)$ as a
function of $\rho$ for different values of $V$
fixing $N=80$ and different $N$ values with fixed $V=8$ respectively. The
density profiles
corresponding to these parameter values are
given in the Figs.~\ref{fig:fig2} and~\ref{fig:fig2a} respectively.
The presence of the incompressible phases, that is, the MI, the CDW-I
and II in the system can be inferred from the \emph{formation of a
peak} in $N(\rho)$ at commensurate densities. For example, existence
of a Mott plateau in the density profile for $V$ ranging between $0$ and $5$ (
see Fig~\ref{fig:fig2}) correlates with a peak in $N(\rho)$ at
$\rho=1$. Similarly peaks in $N(\rho)$ at $\rho=2$ correlate
with the formation of CDW-II phases in the density profile. Similar
conclusions can be drawn from Fig~\ref{fig:fig6a}. Comparing with the
density profile in Fig~\ref{fig:fig2a}, we can conclude that the formation of
peaks in $N(\rho)$ at integer densities can be correlated with the existence of
the solid phases, i.e., MI or CDW. 

In order to distinguish between the two solid phases, i.e, the CDW and MI
phase, we calculate the structure factor, as defined in
Eq.~\ref{eq:sk}. Fig.~\ref{fig:fig7} shows the structure factor in
momentum space as $V$ is varied for $N = 40$, while
Fig.~\ref{fig:fig7a} has fixed $V = 8.0$ for different $N$ values.
From the phase diagram (see the canonical trajectory in
Fig.~\ref{fig:fig1a}) for $N = 40$ the phases at low $V$ values are
the superfluid and the Mott insulator. However, for higher
values of $V$, a CDW-I phase is possible. The CDW oscillations in the
density profile translates to the formation of a peak at $q = \pi$
in the structure factor. As $V$ increases, this peak at $q = \pi$
grows in magnitude reaching its maximum value when the trap center
exhibits the CDW crystalline structure. However this crystalline structure is
possible for a CDW or a SS phase. For example, for $N = 40$ and $V = 8.0$, the
center of the trap is in the supersolid phase that has the CDW-like crystalline
structure and is compressible like a superfluid. Hence the next step is to
distinguish between
the CDW ordered phases that could be either compressible (SS phase) or
incompressible (CDW phase itself). While this can be established locally
with the behavior of compressibility as a function of the distance
from the trap center, a global signature that can be used to check
for the presence of a SS phase in the trap is the momentum distribution
$n(q)$~\cite{scarola-demler-das}.

\bfig[ht]
  \begin{center}
   \includegraphics[angle = 0, width = 3.4in, clip = true]{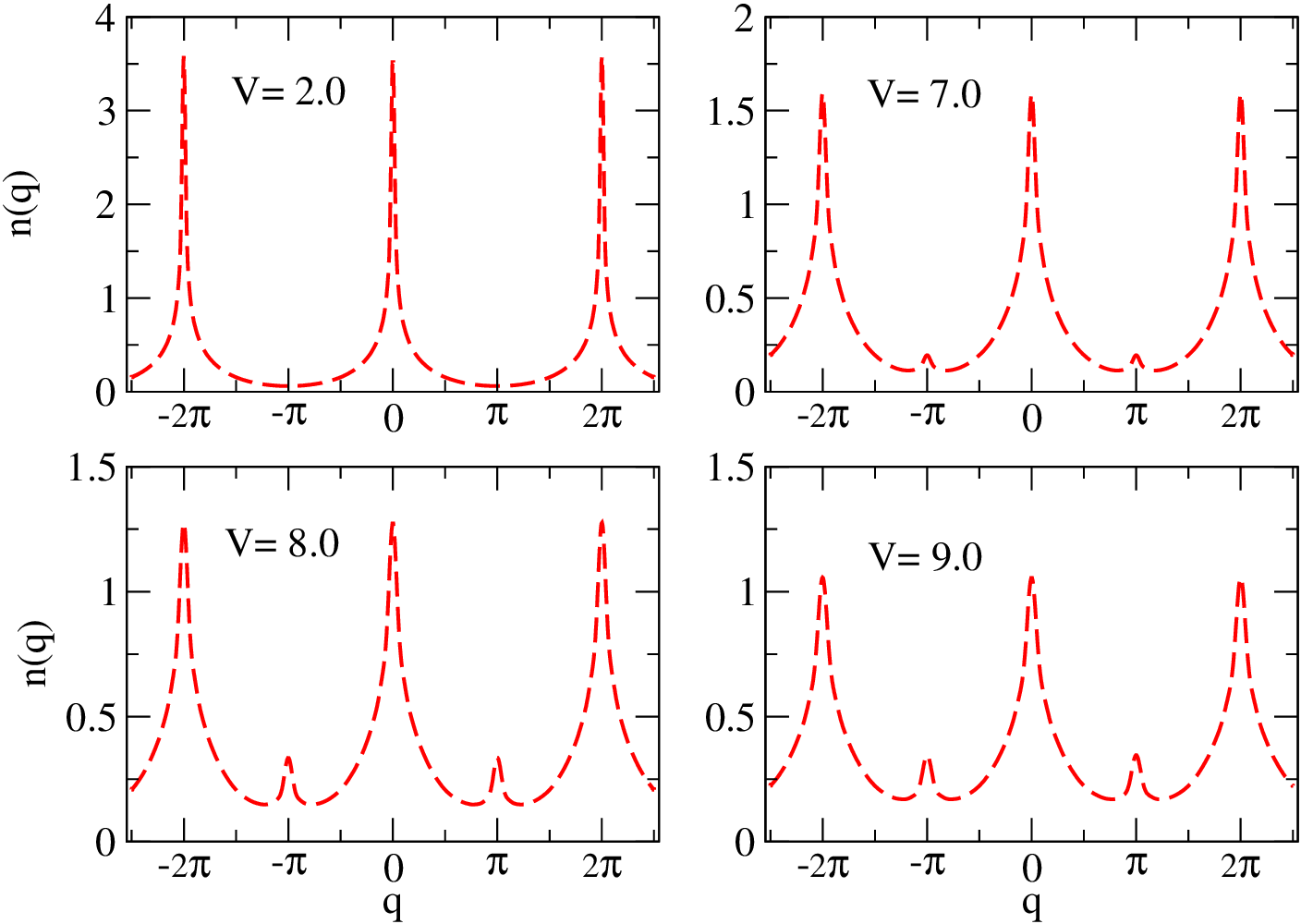}
  \end{center}
  \caption{(Color on-line) Momentum distribution as a function of $q$ for $N = 40$ and different values of $V$.}
  \label{fig:fig8}
  \begin{center}
   \includegraphics[angle = 0, width = 3.4in, clip = true]{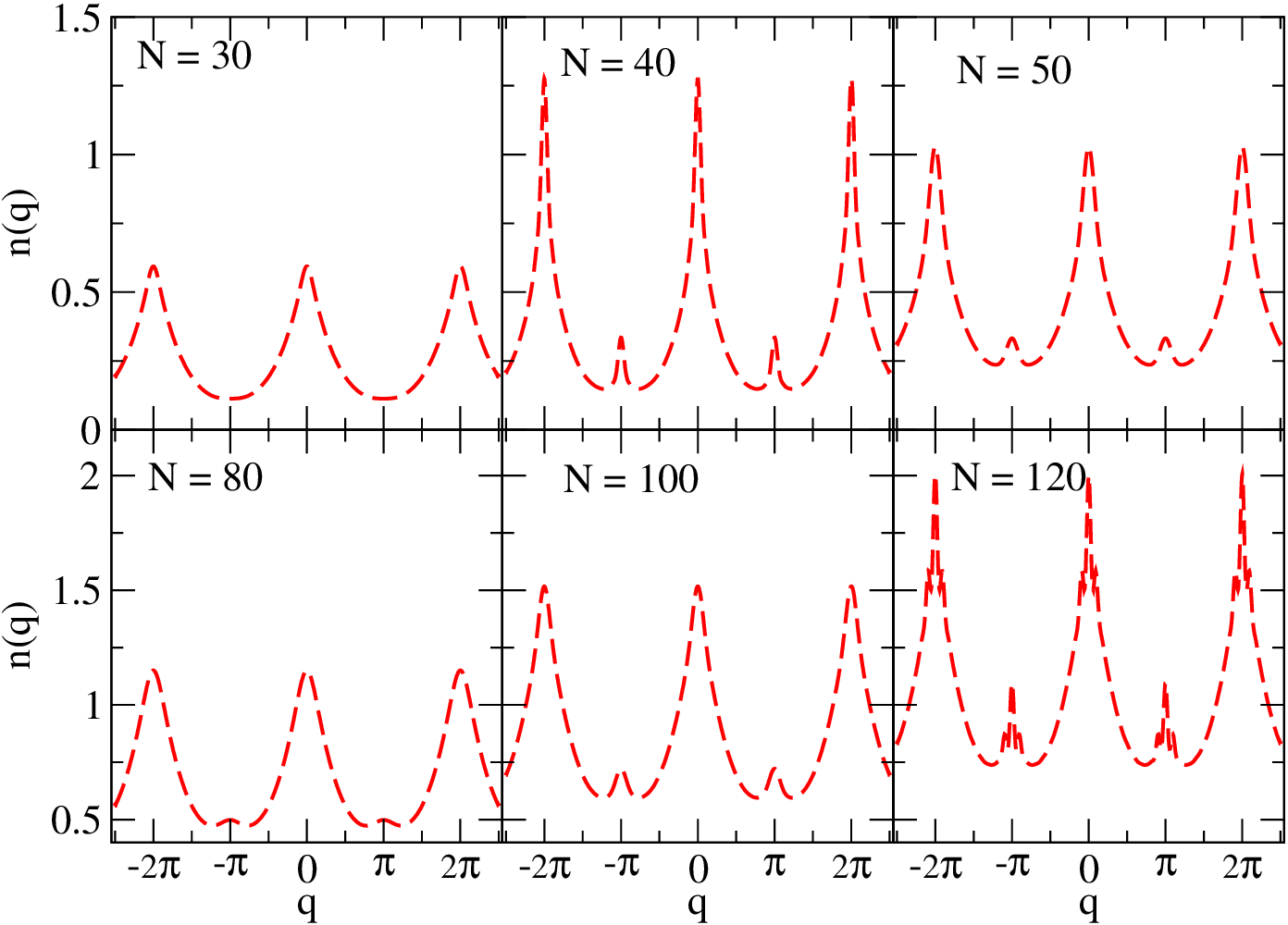}
  \end{center}
  \caption{(Color on-line) Momentum distribution as a function of $q$ for $V = 8.0$ and different values of $N$.}
  \label{fig:fig8a}
\efig

In experiments, the bosons in the optical lattice are allowed to expand and the
interference pattern in the density is recorded. The density distribution is
mirrored in the momentum distribution defined as,
\beq
    n(q) = \D\frac{1}{L} \sum_{k, l = 1}^L \la
a_k^{\dagger} a_l \ra \exp(i q (k - l))
    \label{eqn:mom_dist1}
\eeq
which then provides global information about the various phases present in
the system. Figures~\ref{fig:fig8} and~\ref{fig:fig8a} show the momentum
distribution, respectively, for various values of $V$ with fixed $N = 40$ and for various values of $N$ with fixed $V=8.0$.
We see that in addition to the peaks at $q = 0$ and $q
= \pm 2 \pi$, there are peaks around $q = \pi$. In order to understand the
reason for this peak at $q = \pi$, let us look at the momentum distribution for
the homogeneous  system as in Fig.~\ref{fig:homo_compare}. We choose four densities to demonstrate the features of
the peak in $n(q)$ at $q = \pi$. From the phase
diagram we see that for $V = 8.0$ the homogeneous system with $\rho=0.42$ is in the superfluid phase, $\rho=1/2$ and $1$
are, respectively, in CDW-I and CDW-II phases and for $\rho=0.67$, the system is in the
supersolid phase. From Fig.~\ref{fig:homo_compare} we note that the presence of
a supersolid order in the system is accompanied by a peak in the momentum
distribution at $q = \pi$, which is absent in the SF, CDW-I and CDW-II phases. The structure function for the same set of densities 
as given in Fig.~\ref{fig:homo_compare1} show peak at $q=\pi$ when the system is in CDW-I, CDW-II and SS phases. This confirm that the peak in $n(q)$ at $q=\pi$ is a clear signature of the supersolid phase.

Therefore for the trapped systems, when the phases co-exist, we note that a
peak in the momentum distribution function at $q = \pi$ signals the presence of
a supersolid phase somewhere in the trap, and the peak height being maximum when
the supersolid occupies the center of the trap.
We summarize below the signatures of the different ground state phases for the
inhomogeneous extended Bose-Hubbard model:

\bfig[ht]
  \begin{center}
   \includegraphics[angle = 0, width = 3.4in, clip =
true]{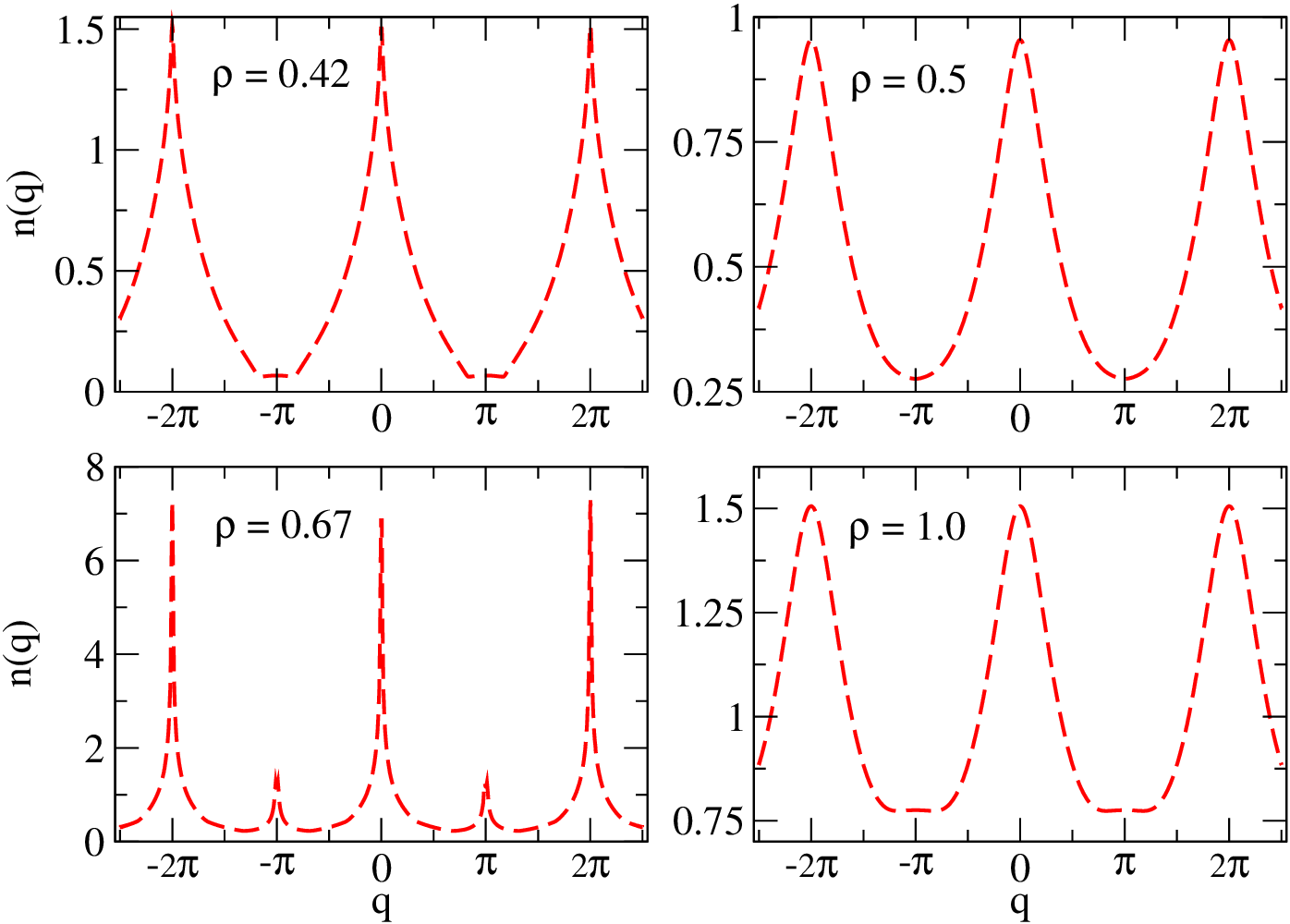}
  \end{center}
  \caption{(Color on line) Momentum distribution for homogeneous case. Note that
when the
system is in the Supersolid phase, a peak at $q = \pi$ develops.}
\label{fig:homo_compare}
    \begin{center}
     \includegraphics[angle = 0, width = 3.4in, clip =
true]{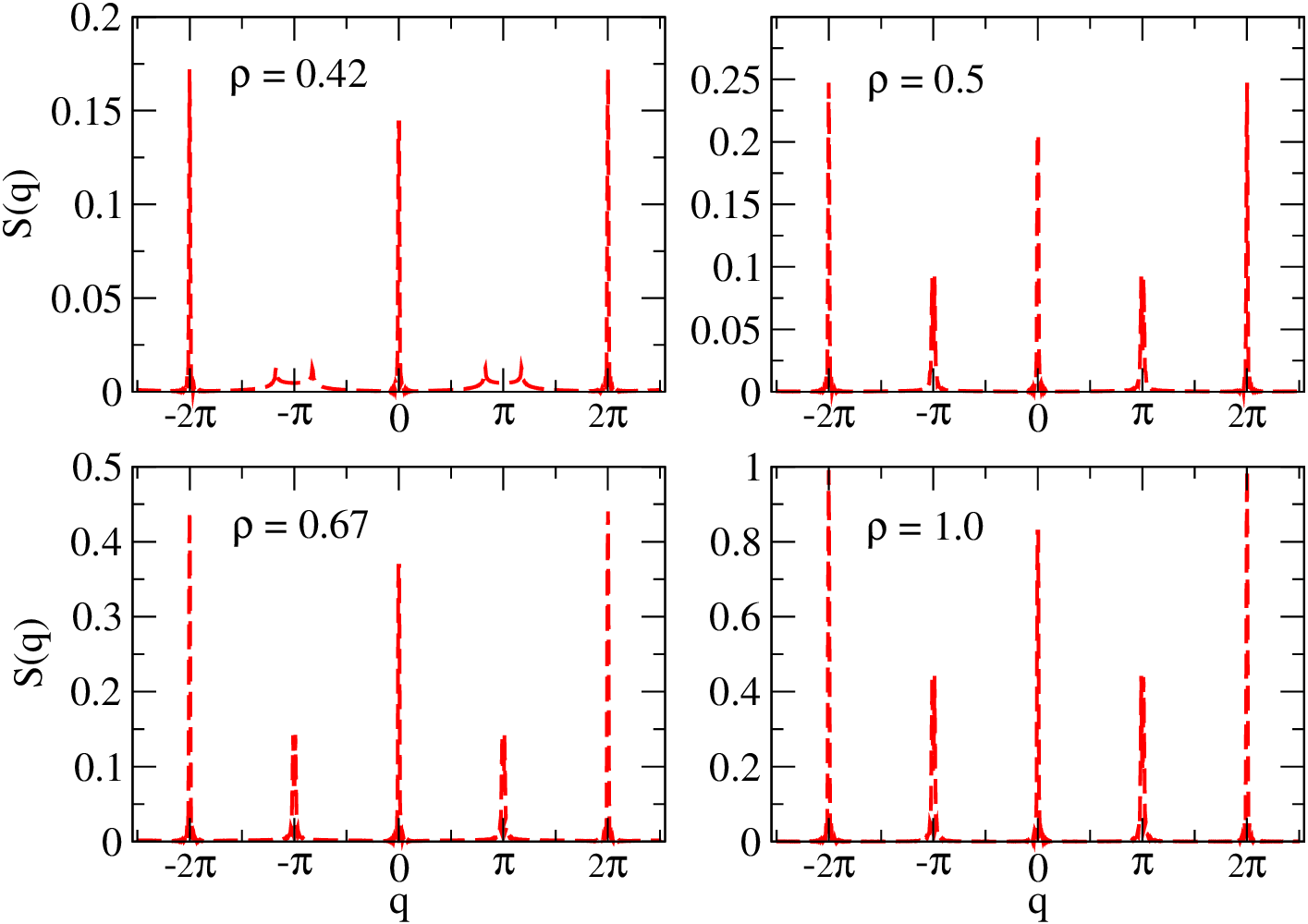}
    \end{center}
    \caption{(Color on line) Structure function for homogeneous case. Note that
when the
system is in the Supersolid phase, a peak at $q = \pi$ develops.}
  \label{fig:homo_compare1}
\efig

\begin{itemize}
    \item MI - Phase:
    \begin{itemize}
        \item Peaks in $N(\rho)$ at integer densities.
                \item No peaks in the momentum distribution at $n(q = \pi)$.
                \item No peaks in the structure function at $S(q = \pi)$.
    \end{itemize}
    \item CDW Phase:
    \begin{itemize}
        \item Peaks in $N(\rho)$ at integer densities.
        \item No peaks in the momentum distribution at $n(q = \pi)$.
        \item Peaks in the structure function at $S(q = \pi)$.
    \end{itemize}
    \item supersolid phase:
    \begin{itemize}
        \item No Peaks in $N(\rho)$ at integer densities.
        \item Peaks in the momentum distribution at $n(q = \pi)$
                \item Peaks in the structure function at $S(q = \pi)$
    \end{itemize}
\end{itemize}

\section{Conclusion}
\label{sect:concl} In conclusion, we have studied a system of
dipolar ultra-cold bosonic atoms in the frame work of the extended
Bose-Hubbard model  in the presence of external harmonic trap. Using
finite size density matrix renormalization group (FS-DMRG) method we
have demonstrated the simultaneous existence of different phases in the
system.  We show the signature of different phases by calculating
different observable quantities such as the on-site number density,
the number fluctuation, the structure factor and the momentum
distribution. We also document global signatures for the ground
phases that can be observed experimentally.

\section{Acknowledgement} R.~V.~P. acknowledges financial support from CSIR and
DST, India.

\end{document}